\documentclass[amsmath,amssymb,aps,prx,reprint,superscriptaddress]{revtex4-2}

\usepackage{amsmath}
\usepackage{xcolor}
\usepackage{amsfonts,amssymb}
\usepackage{graphicx}
\usepackage{hyperref}
\usepackage{academicons}
\usepackage{float}
\usepackage{booktabs} 
\usepackage{multirow,xspace}
\usepackage{orcidlink}

\newcommand{\rrscan}{r$^2$SCAN\xspace}

\begin{document}

\title{Cross-functional transferability in universal machine learning interatomic potentials}

\author{Xu Huang\, \orcidlink{0009-0002-2260-5150}}
\affiliation{Department of Materials Science and Engineering, University of California, Berkeley, California 94720, United States}
\affiliation{Materials Sciences Division, Lawrence Berkeley National Laboratory, California 94720, United States}

\author{Bowen Deng\, \orcidlink{0000-0003-4085-381X}}
\email[]{bowendeng@berkeley.edu}
\affiliation{Department of Materials Science and Engineering, University of California, Berkeley, California 94720, United States}
\affiliation{Materials Sciences Division, Lawrence Berkeley National Laboratory, California 94720, United States}

\author{Peichen Zhong\, \orcidlink{0000-0003-1921-1628}}
\affiliation{Department of Materials Science and Engineering, University of California, Berkeley, California 94720, United States}
\affiliation{Materials Sciences Division, Lawrence Berkeley National Laboratory, California 94720, United States}

\author{Aaron D. Kaplan\, \orcidlink{0000-0003-3439-4856}}
\affiliation{Materials Sciences Division, Lawrence Berkeley National Laboratory, California 94720, United States}

\author{Kristin A. Persson\,\orcidlink{0000-0003-2495-5509}}
\affiliation{Department of Materials Science and Engineering, University of California, Berkeley, California 94720, United States}
\affiliation{Materials Sciences Division, Lawrence Berkeley National Laboratory, California 94720, United States}

\author{Gerbrand Ceder\,\orcidlink{0000-0001-9275-3605}}
\email[]{gceder@berkeley.edu}
\affiliation{Department of Materials Science and Engineering, University of California, Berkeley, California 94720, United States}
\affiliation{Materials Sciences Division, Lawrence Berkeley National Laboratory, California 94720, United States}

\begin{abstract}
The rapid development of universal machine learning interatomic potentials (uMLIPs) has demonstrated the possibility for generalizable learning of the universal potential energy surface. In principle, the accuracy of uMLIPs can be further improved by bridging the model from lower-fidelity datasets to high-fidelity ones. In this work, we analyze the challenge of this transfer learning problem within the CHGNet framework. We show that significant energy scale shifts and poor correlations between GGA and r$^2$SCAN pose challenges to cross-functional data transferability in uMLIPs. By benchmarking different transfer learning approaches on the MP-r$^2$SCAN dataset of 0.24 million structures, we demonstrate the importance of elemental energy referencing in the transfer learning of uMLIPs. By comparing the scaling law with and without the pre-training on a low-fidelity dataset, we show that significant data efficiency can still be achieved through transfer learning, even with a target dataset of sub-million structures. We highlight the importance of proper transfer learning and multi-fidelity learning in creating next-generation uMLIPs on high-fidelity data.
\end{abstract}

\pacs{}
\maketitle

\section{Introduction}
Atomistic simulations provide a powerful framework for predicting and virtual screening of material properties and have led to multiple predictions of interesting functional materials ~\cite{chen2012carbonophosphates, urban2016computational, jain2016computational}. These simulations are enabled by accurate determination of the potential energy surface (PES) as a function of atomic positions, permitting prediction of stability properties, reaction mechanisms, and dynamic behavior~\cite{unke2020high, li2024representing, kopp2023automatic, ock2024gradnav}. \textit{Ab-initio} quantum chemical calculations such as density functional theory (DFT) directly approximate the PES, however, their computational cost scales rapidly with system size, typically, $\sim \mathcal{O}(N_e^3)$ or $\mathcal{O}(N_e \log N_e)$ with $N_e$ the number of \emph{electrons}~\cite{goringe1997dftscale, beck2000real}, and are therefore limited in the length and time scales that can be achieved. To address these limitations, surrogate energy models such as machine-learning interatomic potentials (MLIPs) have been developed to accelerate atomistic simulations while maintaining $\mathcal{O}(N)$ computational efficiency, with $N$ the number of \emph{atoms}~\cite{Zhang_2018_DPMD}.

MLIPs are parametrized to reproduce energies from \textit{ab-initio} quantum mechanical calculations, such as DFT. The total energy of a material system is decomposed and predicted through a learnable mapping of atomic positions and chemical species, where each atom's contribution is determined by its surrounding local atomic configuration within a defined cutoff radius:
\begin{equation}
\hat{E} = \sum_i^n \phi(\{\vec{r}_j\}_i, \{C_j\}_i), ~ \hat{\boldsymbol{f}}_i = - \frac{\partial \hat{E}}{\partial \boldsymbol{r}_i}.
\end{equation}
The learnable function $\phi$ maps the position vectors $\{\vec{r}_j\}_i$ and chemical species $\{C_j\}_i$ of neighboring atoms $j$ to the energy contribution of atom $i$. Forces $\{\hat{\boldsymbol{f}}_i\}$ are derived as the negative gradient of the total energy with respect to atomic coordinates. The choice of design features $\phi$ is crucial for MLIPs to encode the system's physical and chemical properties, such as using equivariant feature encoding~\cite{batzner20223, cheng2024cartesian} and including atomic charge information~\cite{deng2023chgnet, kim2024learning}.

Recently, universal machine-learning interatomic potentials (uMLIPs) trained on millions of DFT calculations demonstrate promising transferability in atomic simulations across diverse chemical spaces. The uMLIPs such as M3GNet~\cite{chen2022universal}, CHGNet~\cite{deng2023chgnet}, MACE-MP-0~\cite{batatia2023foundation}, SevenNet-MF-0~\cite{kim2024data}, and Orb~\cite{neumann2024orb} have been developed from open-source materials databases such as the Materials Project~\cite{jain2013commentary} and Alexandria~\cite{ghahremanpour2018alexandria}. Industry uMLIPs such as GNoME~\cite{merchant2023scaling}, MatterSim~\cite{yang2024mattersim}, and EquiformerV2-OMAT~\cite{barroso2024open} demonstrate improved PES predictability with larger data and model sizes in various downstream materials modeling tasks such as phonon
spectra prediction, phase diagram construction, catalyst screening, and molecular dynamics simulations~\cite{deng2025systematic, yu2024systematic, lan2023adsorbml, chen2025multi, sivak2024discovering}. 

Despite these successes in improving models and data, there remain challenges for further improvements of uMLIPs. One significant issue reported by~\citet{deng2025systematic} shows a consistent underprediction of energies and forces in uMLIPs~\cite{deng2025systematic}, which calls for improved sampling in uMLIP training datasets. The predominant approach to generate uMLIP datasets relies on DFT calculations using generalized gradient approximations (GGAs), limiting uMLIPs to GGA-level accuracy and posing potential challenges for migrating to higher-accuracy functionals like meta-GGAs. Recently, \citet{kaplan2025foundational} released the MatPES dataset that incorporates regularized strongly constrained and appropriately normed (r$^2$SCAN) meta-GGA functional calculations, which opens the possibility for uMLIPs to migrate to high level of theory. See Ref.~\cite{perdew2001ladder} for a definition of GGAs and meta-GGAs and Ref.~\cite{kaplan2023dftreview} for an overview of their well-established limitations in describing crystalline and molecular systems. 

In this work, we discuss the challenges and practical approaches that help better understand the fine-tuning process in uMLIPs, particularly when dealing with multi-fidelity data transferability across different functionals. By showing the correlation between the labels from different levels of theory, we emphasize the importance of training at the right scale through energy referencing when conducting transfer learning.

\section{Observations}
\subsection{Data challenges in existing universal MLIPs}

An essential component in building improved uMLIPs comes from reliable datasets. The current uMLIP datasets applicable to crystalline materials are predominantly composed of GGA and GGA$+U$-level DFT calculations~\cite{chen2022universal, deng2023chgnet, batatia2023foundation, neumann2024orb}. While GGA-based training data is widely available and computationally efficient to generate, several limitations of GGA are known~\cite{perdew1981sie,zhou2004first, goerigk2017} and other functionals are now available~\cite{sun2015strongly, furness2020accurate, heyd2003hybrid}. A widely used method to alleviate some of the self-interaction in GGA is the Hubbard $U$ correction~\cite{anisimov1991band}, which adds an energy correction to localized electron states (e.g., $d$ or $f$ orbitals). The use of $+U$ is particularly important when dealing with metal oxidation/reduction in formation enthalpies, reaction energies, or electrochemical potentials~\cite{wang2006oxidation, zhou2004first}. At the same time, the application of $+U$ is not appropriate for metallic systems where electron delocalization is appropriate. Because of these conflicting requirements, compatibility schemes between GGA and GGA$+U$ have been designed~\cite{jain2011formation} and some datasets contain a mixture of GGA and GGA$+U$ calculations. We call attention to three data challenges in existing uMLIPs, which were primarily trained with a mixture of GGA/GGA$+U$ DFT calculations.

1. GGA/GGA$+U$ exhibit lower transferability across chemical bonding environments~\cite{goerigk2017}. The Perdew–Burke–Ernzerhof (PBE) GGA~\cite{perdew1996generalized} is found to have a mean absolute error (MAE) of 194 meV/atom dominated by the large error in oxides and strongly bound systems, in a large-scale test on the formation energy of 987 compounds~\cite{kothakonda2022testing}. In contrast, the SCAN meta-GGA functional developed by~\citet{sun2015strongly} predicts formation energies with an MAE of 84 meV/atom.~\citet{isaacs2018performance} also demonstrate that SCAN is more accurate in predicting formation energy for strongly bound compounds, crystal volumes, magnetism, and band gaps, as compared to the PBE GGA. The r$^2$SCAN~\cite{furness2020accurate} revision of the SCAN meta-GGA balances numerical stability with high general accuracy~\cite{kothakonda2022testing} and has therefore become the preferred method to evaluate thermophysical properties of materials~\cite{kingsbury2022performance,kothakonda2022testing, liu2024assessing}. While the demonstrated prediction errors in Ref.~\cite{kothakonda2022testing} are high, it is worth noting that many of the compounds included have formation reactions from molecular species such as H$_2$, N$_2$, O$_2$, and thereby are more similar to cohesive energies. When evaluating only solid-state reactions, energy errors are typically smaller for GGA~\cite{hautier2012accuracy}.

2. The application of the Hubbard $U$ correction to mitigate self-interaction errors in GGA is inherently semi-empirical and non-universal. GGA$+U$ fails to predict accurate energy differences between some compounds with localized electronic states and those with delocalized electronic states~\cite{jain2011formation}.
There is also no precise definition of an ``optimal'' $U$, and approaches such as the linear response method~\cite{cococcioni2005} suggest that such an optimal $U$ would be system-dependent. However, the GGA/GGA$+U$ uMLIP datasets were generated using the same $U$ value for each element regardless of the local environment or formal valence state, calibrated to minimize discrepancies between DFT-calculated oxidation energies and experimental measurements for a limited number of 3$d$ transition metal oxides~\cite{wang2006oxidation,jain2011formation}.

3. To correct for some of the self-interaction error in GGA which is particularly large when calculating the energy of reactions that reflect charge transfer such as oxide formation enthalpies, an \textit{ad hoc} scheme of mixing GGA and GGA$+U$ calculations is typically used to bridge the gap between GGA and GGA$+U$~\cite{jain2011formation, kingsbury2022flexible}. Such coarse-grained, non-universal adjustments can potentially cause issues when fitting a uMLIP, such as sudden jumps of potential energy at the scale of a few hundred meV per atom when moving between training data computed with these mixing schemes.
Last, there is no corresponding mixing scheme applied to the GGA/GGA$+U$ interatomic forces and stresses.
This may be less of an issue as both are derivative properties of a given functional, and thus should be independent of the energy scale of the underlying DFT approximation.
However, this has not been formally verified.

Overall, the use of approximate exchange-correlation functionals, combined with the non-universality of Hubbard $U$ corrections and compatibility adjustments, leads to less accurate and somewhat noisy data within the GGA/GGA$+U$ framework. Such data noise makes it challenging for graph neural network models (GNNs) to accurately learn and capture the underlying interactions within materials. 

\subsection{Cross-functional transferability challenges in universal MLIPs}

One possible solution to overcome the challenges of GGA and GGA$+U$ is to shift the uMLIP training and benchmarking dataset to DFT calculations performed with higher-fidelity functionals. These higher-fidelity calculations come with higher computational costs, leading to challenges in constructing datasets on a substantial scale. One possible solution is to leverage existing lower-fidelity GGA and GGA$+U$ calculations and existing pre-trained uMLIPs as a starting point. 

There are three main strategies to achieve explicit or implicit transferability between multi-fidelity DFT datasets: transfer learning, multi-fidelity learning, and mixed multi-fidelity training. 
\begin{enumerate}
     \item \textbf{Transfer learning} (TL) involves pre-training a large neural network on extensive lower-fidelity datasets. The pretrained weights from this network are then transferred to initialize machine-learning tasks on smaller, higher-fidelity datasets. This approach is both computationally efficient and data-efficient~\cite{hoffmann2023transfer, chen2023data}. However, if the correlation between the two different fidelity datasets is not strong enough, TL is not effective and can even deteriorate the learning performance, known as negative transfer~\cite{wang2019characterizing}.
     \item \textbf{Multi-fidelity learning} can be conducted either at the feature (input) level or at the label (output) level~\cite{gong2022calibrating}, i.e., low-fidelity data is utilized as input features to predict high-fidelity data, or the task of learning high-fidelity data can be transformed into learning the difference between high-fidelity and low-fidelity data, an approach known as \(\Delta\)-machine learning~\cite{ramakrishnan2015big}. Multi-fidelity learning tends to be more computationally expensive than TL~\cite{dral2023learning}. When applying multi-fidelity learned models to make real predictions for unknown cases, one must first calculate low-fidelity data to obtain input features (input level) or use it to add the predicted difference to get the final high-fidelity prediction (output level).
     \item \textbf{Mixed multi-fidelity training} aims to simultaneously learn and predict datasets of varying fidelity levels.~\citet{chen2021learning} encoded the fidelity of each dataset and embedded the dataset type as a vector in the global state feature input to the M3GNet model for band gap prediction.~\citet{ko2025data} adopted this method to construct highly accurate GNN-based interatomic potentials for two model systems---silicon and water.~\citet{allen2024learning} used meta-learning techniques to build pre-trained potentials that simultaneously incorporate information from multiple large organic datasets, calculated at different levels of theory.~\citet{kim2024data} developed a high-fidelity MLIP by one-hot encoding each fidelity, concatenating it to the scalar part of the input node feature at each linear layer, and adding different atomic energy shift scale blocks for each fidelity database to the SevenNet model. Similar to TL, mixed-fidelity training tends to be computationally expensive when additional poorly correlated data are added to the trained model.
 \end{enumerate}
 
Each of the three strategies presents its own advantages and challenges. So far, no clear evidence exists that TL consistently outperforms multi-fidelity learning or mixed multi-fidelity approaches, or vice versa. In this work, we focus on how to tackle the transferability challenges of efficient TL across GGA/GGA$+U$ mixed data and r$^2$SCAN data in the CHGNet model, though our conclusion should hold more generally for other uMLIPs.

\section{Results}

In this section, we use a r$^2$SCAN dataset, MP-r$^2$SCAN, parsed from Materials Project~\cite{jain2013commentary} r$^2$SCAN relaxation trajectories, for high-fidelity training tasks. Following the data parsing criteria described in \hyperref[subsec:data]{Data preparation}, we obtain 34,927 material IDs with 238,247 structures. Compared to the MPtrj Dataset~\cite{deng2023chgnet}, which has 145,923 materials IDs with 1,580,395 structures, the MP-r$^2$SCAN is significantly smaller in size.

\begin{figure*}[htb]
\centering
\includegraphics[width=1.0\linewidth]{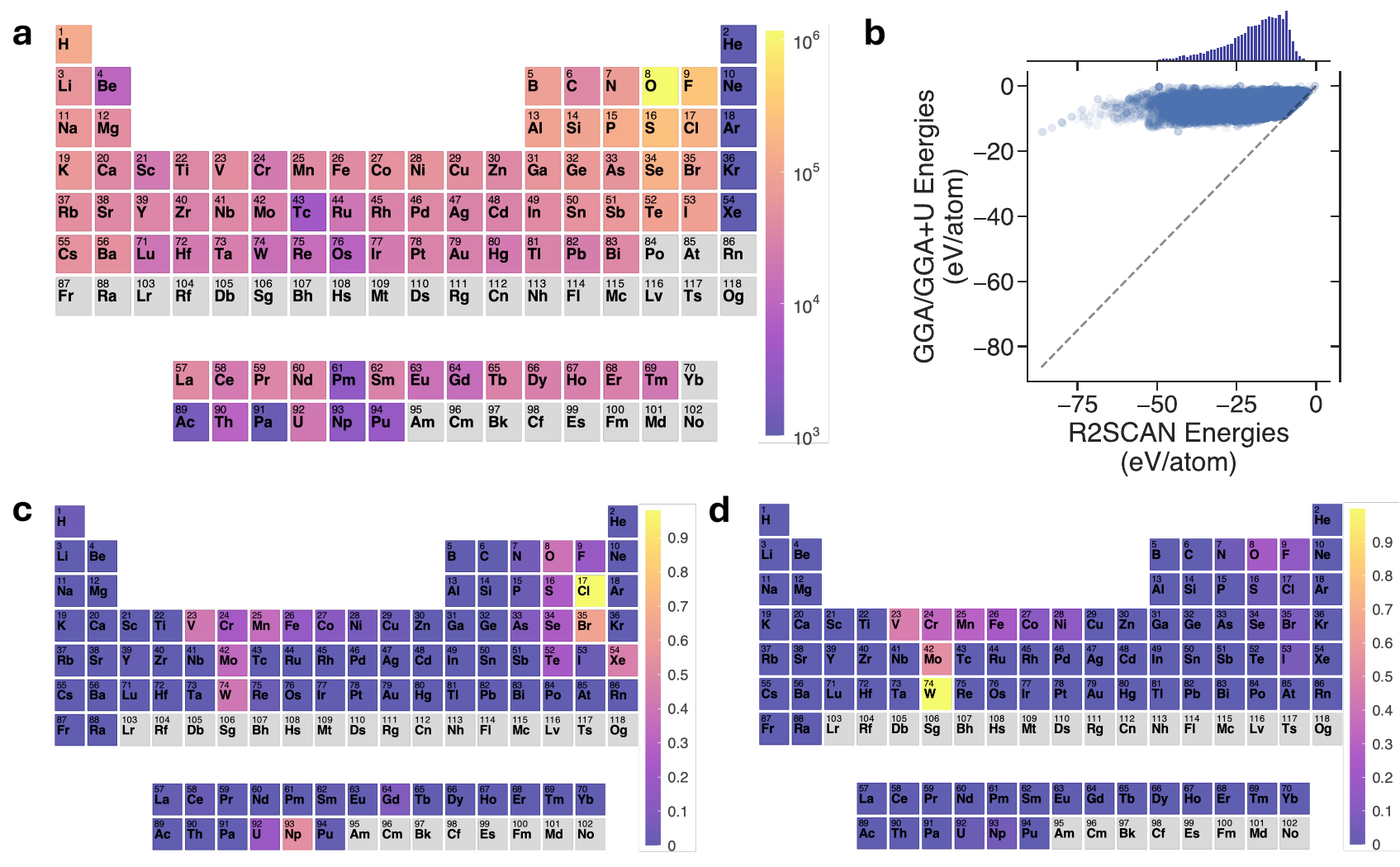}\caption{\textbf{Statistical analysis of the energy data.} 
\textbf{a} Element distribution of the MP-r$^2$SCAN dataset of 238,247 structures. The color indicates the total number of occurrences of an element in the MP-r$^2$SCAN dataset with a lower cutoff of 1000. \textbf{b} Total Energy of materials computed from GGA/GGA+U vs. r$^2$SCAN functionals. Each point represents a material with a materials ID that has r$^2$SCAN calculations in Materials Project, with the $x$-axis showing the total energy after r$^2$SCAN structure relaxation and the $y$-axis showing the total energy after GGA/GGA+U structure relaxation. The marginal histograms on the top and right illustrate the distributions of total energies for the same collection of materials, as calculated by r$^2$SCAN and GGA/GGA$+U$, respectively. \textbf{c--d} Feature importance in the formation energy differences between GGA/GGA$+U$ mixing and r$^2$SCAN. Each element is treated as a feature, with its importance indicated by colors on the periodic table. Higher values correspond to greater importance and therefore larger energy difference between GGA/GGA$+U$ and r$^2$SCAN.
Panel \textbf{c} presents the feature importance when anion and compatibility corrections are included in the mixed GGA/GGA$+U$ data, and panel \textbf{d} presents the feature importance without these adjustments.
Compositional corrections are applied primarily to pnictogens, chalcogens, and halogens.}
\label{data_analysis}
\end{figure*}

Figure~\ref{data_analysis}a presents the element distribution in the MP-r$^2$SCAN dataset with a total of 238,247 structures. The color of each element indicates the total number of times each element is present in the MP-r$^2$SCAN dataset, with a lower cutoff of 1000. Elements with 1000 or fewer occurrences all share the same color. The MP-r$^2$SCAN dataset covers 88 elements in the periodic table.

\subsection{Energy differences across two functionals}
\label{subsec:functional_difference}
Machine learning transferability can be quantified by assessing the correlations between the source and target datasets~\cite{gerace2022probing}.
To investigate the feasibility and effectiveness of TL between DFT functionals, we analyze the scale of the total energy differences between r$^2$SCAN and GGA/GGA$+U$.

Figure~\ref{data_analysis}b presents the comparison of the relaxed total energies calculated using r$^2$SCAN ($x$-axis) and GGA/GGA$+U$ ($y$-axis), which represent the training label of most uMLIPs. In Fig.~\ref{data_analysis}b, each point represents a single compound from the Materials Project, and the corresponding GGA/GGA$+U$ energies have applied anion and compatibility corrections~\cite{wang2021framework}. The marginal histograms on the top and right side show the distributions of energies calculated using r$^2$SCAN and GGA/GGA$+U$, respectively, for all r$^2$SCAN materials IDs
in Materials Project. As depicted in Fig.~\ref{data_analysis}b, the total energy of r$^2$SCAN and GGA/GGA$+U$ are distributed on different scales. The shift from the GGA/GGA$+U$ to r$^2$SCAN is at the scale of 0--70\, \text{eV/atom}, which is significantly larger than the energy accuracy of uMLIPs (\(\sim 30\  \text{meV/atom}\)), indicating these r$^2$SCAN energy labels are not directly transferrable without proper reference or normalization. 

These eV/atom scale energy shifts between functionals are related to the ambiguity in the Kohn-Sham energy levels which have an arbitrary reference energy~\cite{Choe_West_Zhang_2021, Ihm_Zunger_Cohen_2001, how2025adaptive}. 
These energy shifts are well understood in electronic structure theory and do not contribute to any physical quantities due to the cancellation of energy references in any physical property. 
The total energy itself is not a physically measurable quantity, as it is ``gauge dependent'' on the vacuum level, but energy differences such as the cohesive energy are measurable and gauge invariant\cite{kittel2018introduction}.
Because MLIPs are typically trained on absolute total energies, these eV/atom scale energy differences from GGA/GGA$+U$ and r$^2$SCAN can cause significant challenges in TL.

One method to remove the significant total energy shifts is by fitting the MLIPs with physical quantities such as formation energies, which has been shown to be easier to transfer in crystal graph attention networks~\cite{hoffmann2023transfer, schmidt2021crystal}. The formation energies describe the strengths of the interactions that form the compound from pure elemental phases and are better correlated between different functionals than the total energy labels, although small deviations can still be present due to the different levels of accuracy. 

To determine which elements contribute most to the formation energy differences between r$^2$SCAN and GGA/GGA$+U$ calculations, we queried the formation energies from Materials Project and fitted decision tree models on the formation energy differences through \texttt{scikit-learn}~\cite{Quinlan_1986_decisiontree}. The input to this model is the compositional fraction matrix of all materials with r$^2$SCAN materials IDs
in Materials Project, and the target variable is the formation energy difference between the two functionals. We calculated the feature importance (see \hyperref[subsec:fea_impor]{Feature importance}) for each element and plotted the strength of the importance through the color bar in the periodic table in Fig.~\ref{data_analysis}c and d. The importance of a feature is computed as the normalized total reduction of the criterion brought by that feature. The higher the value the more important the feature. Figure~\ref{data_analysis}c presents the feature importance with GGA/GGA$+U$ mixing and anion corrections included, and Fig.~\ref{data_analysis}d includes the same analysis but with \emph{uncorrected} GGA/GGA$+U$ formation energies.

In Fig.~\ref{data_analysis}c, we observe that $d$-block elements such as V, Cr, Mn, Fe, Co, Ni, Mo, and W exhibit high importance, indicating they significantly contribute to the formation energy differences between GGA/GGA$+U$ and r$^2$SCAN. These are precisely the elements for which Hubbard $U$ corrections and compatibility adjustments are applied in transition metal oxides and fluorides. Similarly, $p$-block elements with high importance---O, F, S, Cl, Se, Br, and Te---also undergo compatibility corrections when they serve as anions in compounds. Notably, Cl exhibits a very high feature importance. We can attribute the relatively higher feature importance of Cl to two sources: (i) the compatibility scheme imposed on GGA/GGA$+U$ energies places the second largest correction ($-0.614$ eV/atom in magnitude) to Cl, second only to oxides ($-0.687$ eV/atom in magnitude); (ii) PBE struggles to describe the weaker covalency and van der Waals interactions typical of ionic crystals~\cite{sun2013alpha}, whereas \rrscan describes both covalent and ionic bonding reasonably well~\cite{furness2020accurate} and improves the description of medium-range van der Waals interactions~\cite{yang2019rationalizing,ning2022r2scan_rvv10}. The differences in Fig.~\ref{data_analysis}c and d show clearly that the removal of the corrections scheme almost eliminates the higher feature importance of the chalcogens and halogens seen in Fig.~\ref{data_analysis}c. Without the energy correction scheme, the eight transition metals, O, and F remain a higher feature importance (see Fig.~\ref{data_analysis}d).

\subsection{TL with different atomic reference energies}
\label{approach}

\begin{figure}[htb]
\centering
\includegraphics[width=1.05\linewidth]{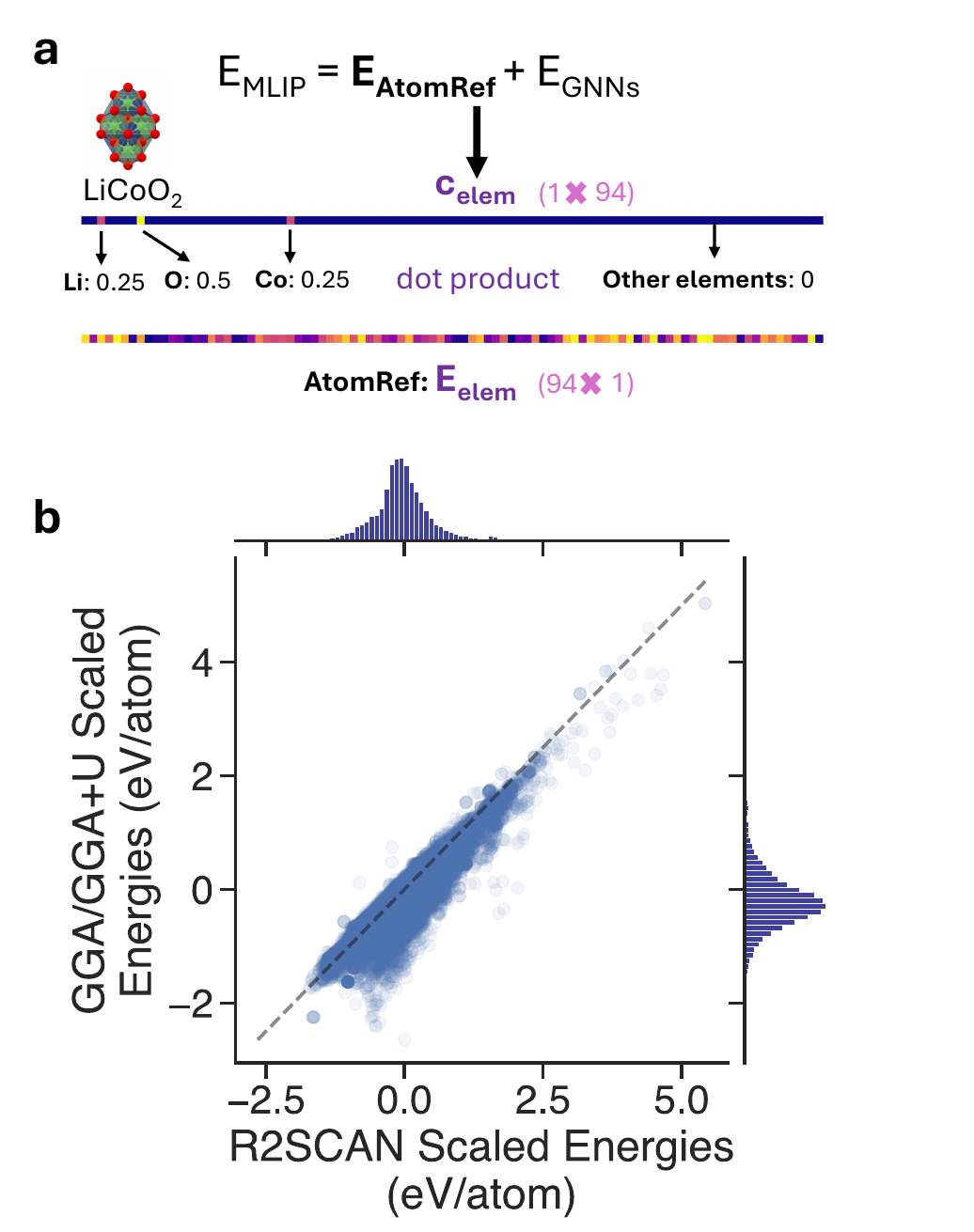}
\caption{\textbf{Illustration of AtomRef and correlation improvement through scaled energies.} 
\textbf{a} Schematic representation of the role and application of AtomRef in calculating total energies. The energy contribution from AtomRef is obtained by taking the dot product of the composition row vector (with LiCoO$_2$ used here as an example) and the AtomRef vector. 
\textbf{b} The correlation between the scaled energies of GGA/GGA$+U$ and r$^2$SCAN (total energies with the respective AtomRefs subtracted). The marginal histograms on the top and right illustrate the distributions of r$^2$SCAN and GGA/GGA$+U$ scaled energies, respectively, for the same collection of materials.}
\label{correlation}
\end{figure}

Shifting the PES with a constant value for each element is an effective and commonly used approach in training GNN-based MLIPs. As described in Fig.~\ref{correlation}a, in CHGNet and other models like M3GNet, NequIP~\cite{batzner20223} and CACE~\cite{cheng2024cartesian}, the prediction of total energies (per atom) is divided into two parts: $E_{\text{AtomRef}}$ and $E_{\text{GNNs}}$~\cite{chen2022universal}. First, the composition row vector $\mathbf{c}_{\text{elem}}$ and atomic reference energies (AtomRef) $\mathbf{E}_{\text{elem}}$ are obtained, and their dot product gives $E_{\text{AtomRef}}$. The composition vector $\mathbf{c}_{\text{elem}}$ represents the fraction of each element in the structure, and in CHGNet, its dimension is $1 \times 94$. Next, a composition model is used to fit a linear regression of total energies, where $\mathbf{E}_{\text{elem}}$ are the weights: \: \begin{equation}
    \mathbf{E}_{\text{elem}} = (\mathbf{A}^T \mathbf{A})^{-1} \mathbf{A}^T \mathbf{E}_{\text{total}}
\end{equation}
Here, $\mathbf{A}$ is the composition matrix obtained by stacking $\mathbf{c}_{\text{elem}}$ for all structures in the training set, and $\mathbf{E}_{\text{total}}$ is the matrix of total energies. Subsequently, the remaining fine-grained energy is predicted by GNNs. Overall, the total energy prediction of a structure can be expressed using $
    E_{\text{total}} =\mathbf{c}_{\text{elem}} \cdot \mathbf{E}_{\text{elem}} + 
    E_{\text{GNNs}}$. Both AtomRef, which represent the weights of the composition model, and GNNs can be trainable.

For cross-functional TL on a uMLIP with a fitted AtomRef from GGA/GGA$+U$ total energies, one can refit the uMLIP's AtomRef to shift the uMLIP's energy to the scale of new DFT labels and, in principle, improve the correlation between pre-training and fine-tuning datasets. Refitting the AtomRef essentially replaces the fitted GGA/GGA$+U$ AtomRef with the fitted r$^2$SCAN AtomRef and shifts the uMLIP's predicted energy scale to r$^2$SCAN. Figure~\ref{correlation}b shows that, after replacing the AtomRef, a stronger correlation between GGA/GGA$+U$ and r$^2$SCAN total energies can be achieved. 

Indeed, the Pearson's correlation coefficient $\rho$ improves from 0.0917 between the unmodified GGA/GGA$+U$ and r$^2$SCAN datasets to 0.9250 between the r$^2$SCAN energies (with r$^2$SCAN AtomRef subtracted) and the GGA/GGA$+U$ energies (with GGA/GGA$+U$ AtomRef subtracted). 

\begin{table*}[t]
  \centering
  \setlength{\tabcolsep}{3pt}
  \renewcommand{\arraystretch}{1.2} 
  \begin{tabular}{lcccccc}
  \toprule
        Methods & Energy MAE & Force MAE & Stress MAE & Magmom MAE & Decomposition energy MAE & Formation energy MAE \\
         & (meV/atom) & (meV/\text{\AA}) & (GPa) & ($\mu_B$) & (meV/atom) & (meV/atom) \\
  \midrule 
         Method 1 & 27 & 45 & 0.239 & \textbf{0.019} & 37.44 & 43.11 \\
         Method 2 & 26 & 54 & 0.266 & 0.027 & 41.22 & 52.43 \\
         Method 3 & 26 & 52 & 0.257 & 0.026 & 38.54 & 39.78 \\
         Method 4 & \textbf{17} & \textbf{38} & \textbf{167} & 0.023 & \textbf{23.66} & \textbf{29.38} \\
  \bottomrule
  \end{tabular}
  \caption{Energy, force, stress, magnetic moment (magmom), decomposition energy, and formation energy prediction mean absolute errors (MAEs) of different methods. Method 1: Training from scratch; Method 2: TL with trainable AtomRef; Method 3: TL with frozen AtomRef; Method 4: TL with r$^2$SCAN AtomRef.}
  \label{efsm_decomp_maes}
\end{table*}

To compare in more detail how well various strategies for aligning energies from different functionals perform, we performed an ablation study using four training strategies to either pre-train or fine-tune CHGNet on the MP-r$^2$SCAN dataset.

\begin{itemize}
    \item 
    \textbf{Method 1: Training from scratch.} We first fitted AtomRef using the r$^2$SCAN total energies, randomly initialized the GNN parameters of CHGNet, and then trained the GNNs on the MP-r$^2$SCAN dataset while keeping the r$^2$SCAN AtomRef frozen.
    
    \item
    \textbf{Method 2: TL with trainable AtomRef.} Starting from the GGA/GGA$+U$-pre-trained CHGNet, both the GNN parameters and the AtomRef were allowed to be trainable during TL. In this manner, the AtomRef, initially set to the fitted GGA/GGA$+U$ AtomRef, was gradually updated throughout the TL process.
    
    \item
    \textbf{Method 3: TL with frozen AtomRef.} Again using the GGA/GGA$+U$-pre-trained CHGNet as the starting point, only the GNN parameters were allowed to be trainable during TL. As a result, the AtomRef remained fixed at the fitted GGA/GGA$+U$ AtomRef, forcing the GNNs to transfer and accommodate to the large energy differences observed in Fig. \ref{data_analysis}b.
    
    \item
    \textbf{Method 4: TL with r$^2$SCAN AtomRef.} We first replaced the GGA/GGA$+U$ AtomRef in the pre-trained CHGNet model with the r$^2$SCAN AtomRef, and then performed TL on the GNNs while keeping the r$^2$SCAN AtomRef frozen.
\end{itemize}
Table~\ref{efsm_decomp_maes} presents the MAEs on the test set for energy, force, stress, and magnetic moment (magmom) predictions (see \hyperref[subsec:data]{Data preparation} for details on data splitting). Methods 2 and 3 (TL with trainable and frozen AtomRef, respectively) yield similar performance across all metrics, with Method 1 (Training from scratch) achieving a comparable energy error (27 meV/atom) but reduced force (45 meV/\AA) and stress error (0.239 GPa). This suggests that without properly shifting the reference energy, neither Method 2 nor Method 3 benefits from the GGA/GGA$+U$ pre-training. In contrast, Method 4 (TL with r$^2$SCAN AtomRef) attained the lowest MAEs for energy, force, and stress, indicating that the optimal approach to fine-tuning MLIPs is to first shift the reference energy and then train the GNNs.

\begin{figure*}[htb]
\centering
\includegraphics[width=0.83\linewidth]{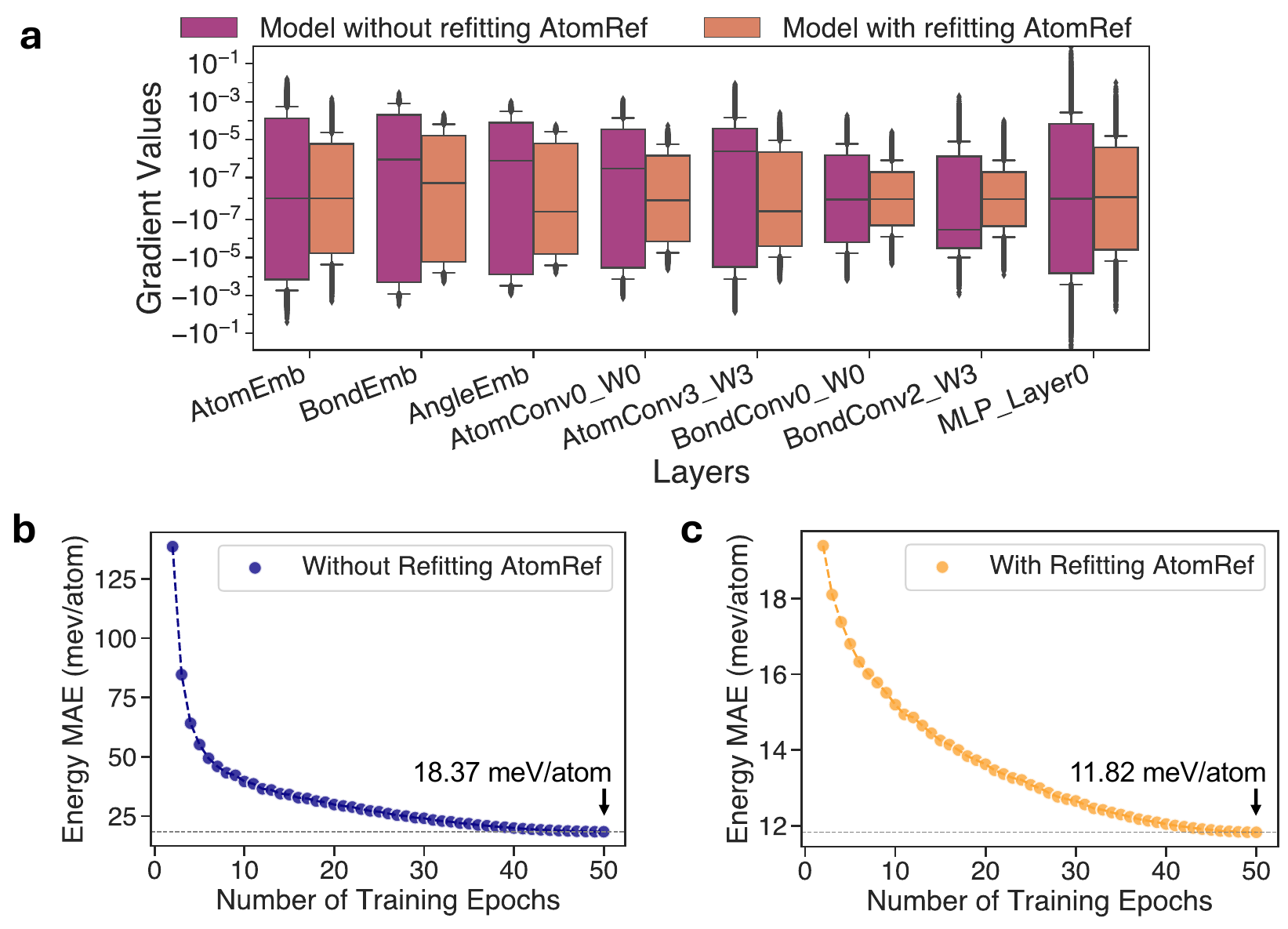}
\caption{\textbf{Comparison of the model's training performance with and without AtomRef refitting.} 
\textbf{a} Gradient values recorded every 1/10 of an epoch for various model layers during the first transfer learning epoch, comparing models with and without AtomRef refitting. The layers include ``AtomEmb'' (atom embedding), ``BondEmb'' (bond embedding), ``AngleEmb'' (angle embedding), ``AtomConv0\_W0'' and ``AtomConv3\_W3'' (weights of the two-body atom convolution layers), ``BondConv0\_W0'' and ``BondConv2\_W3'' (weights of the two-body bond convolution layers), and ``MLP\_Layer0'' (weights of the first layer in the multi-layer perceptron). 
\textbf{b} Energy training history for Method 3, showing the lowest energy MAE of 18.37 meV/atom at the last epoch. 
\textbf{c} Energy training history for Method 4, showing the lowest energy MAE of 11.82 meV/atom at the last epoch.}
\label{training}
\end{figure*}

Figure~\ref{training} shows the model training gradients and training errors vs. epochs for Method 3 and Method 4 during the TL. Figure~\ref{training}a illustrates the range of gradient values for several representative model layers. Gradient
values are recorded every 1/10 of an epoch for these model layers during the first transfer learning epoch. We observe that Method 3 without refitting AtomRef exhibits gradient magnitudes at least one order larger than those of Method 4 with refitting. Figures~\ref{training}b and~\ref{training}c show the evolution of energy MAE during the full training process of 50 epochs, without and with AtomRef adjustments, respectively. Figure~\ref{training}b displays larger initial and final energy MAE, indicating a less effective training process. In contrast, Figure~\ref{training}c demonstrates that refitting AtomRef results in a more stable and reliable training history.

\subsection{Stability prediction from MLIPs}

As a more stringent prediction test, we evaluate relative stability of compounds through the convex hull construction. Relative stability of a compound can be measured by its decomposition energy, calculated by the total energy difference between a given compound and its competing compounds in a specific chemical space. This is a more stringent test than measuring MAE, as the scale of decomposition energy is small and relies on significant error cancellation in DFT~\cite{bartel2020critical}.
\begin{figure*}[htb]
\centering
\includegraphics[width=0.9\linewidth]{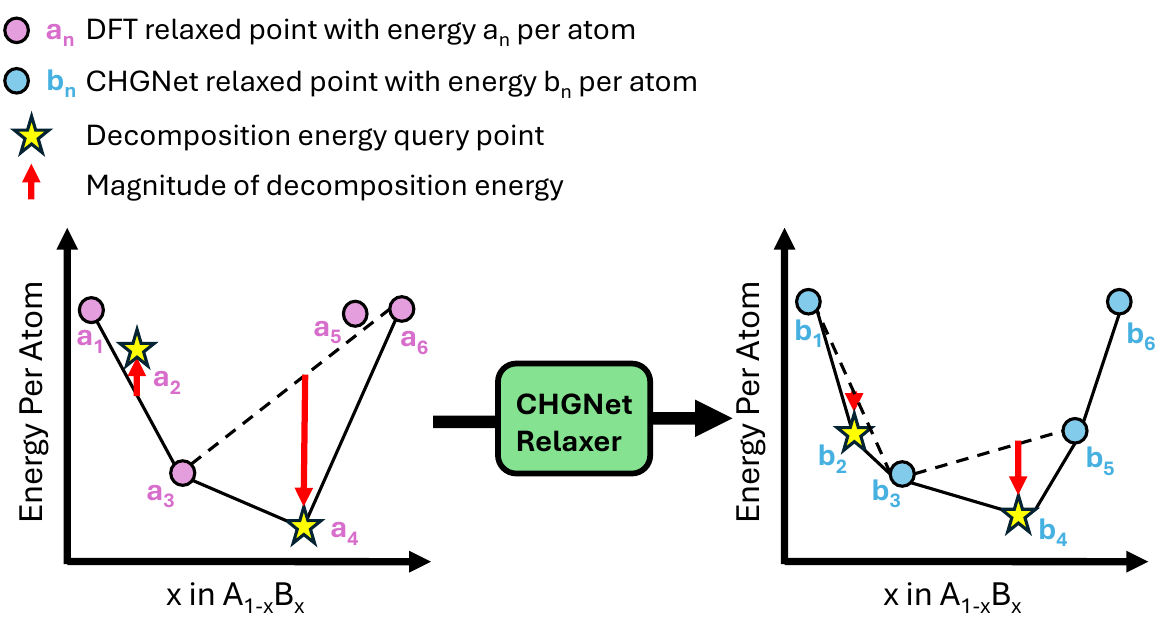}
\caption{\textbf{Decomposition energy prediction workflow.} The left plot shows a schematic of a convex hull energy diagram constructed using r$^2$SCAN DFT-calculated data, providing decomposition energy values based on competing phases identified in the DFT phase diagram (e.g., for a$_2$, the competing phases are a$_1$ and a$_3$; for a$_4$, they are a$_3$ and a$_6$). The right plot schematically shows the convex hull constructed by CHGNet-relaxed energies. The decomposition energy and model-identified competing phases differ from DFT.}
\label{decomp_workflow}
\end{figure*}

Figure~\ref{decomp_workflow} presents the general workflow for predicting decomposition energy. Predicting decomposition energy with uMLIPs is particularly challenging as it depends not only on the energy of a single material but also on that of the neighboring competing phases in a phase diagram~\cite{bartel2022review}. The physical outcome of decomposition energy is binary with negative values indicating stable compounds and positive values indicating unstable or metastable compounds. As such, small non-systematic energy errors from MLIPs will easily alter the stable entries in the phase diagram, by changing the decomposition energy from small negative values to positive values and vice versa. This issue is further exacerbated by the fact that machine learning models exhibit poorer error cancellation compared to DFT~\cite{bartel2020critical}. 

We constructed all phase diagrams in the chemical space of our dataset using r$^2$SCAN DFT data and calculated the decomposition energy as the ground truth. A similar phase diagram can be constructed by the fine-tuned CHGNet, which allows the determination of CHGNet predicted decomposition energy. The initial configurations for all structures are sourced from Materials Project and further relaxed using the pre-trained or fine-tuned CHGNet models of corresponding methods. This process relies solely on the uMLIP's capability to obtain relaxed energies and relative stabilities between polymorphs, without requiring additional information from the DFT phase diagram.

Table~\ref{efsm_decomp_maes} also presents benchmark results for the decomposition energy prediction MAEs of four methods on the MP-r$^2$SCAN test set (see \hyperref[subsec:data]{Data preparation} for data splitting). The MAEs of Methods 2 and 3 (41.22 and 38.54 meV/atom, respectively) are slightly larger than that of Method 1 (37.44 meV/atom), again indicating no benefit from conventional TL methods. In contrast, Method 4, which uses r$^2$SCAN-specific AtomRef, achieves an MAE of 23.66 meV/atom, at least 13.5 meV/atom lower than the others. Additionally, Table~\ref{efsm_decomp_maes} shows the formation energy MAEs for the pre-trained or fine-tuned CHGNet models, where formation energy is defined as the energy difference between a compound and its constituent elements in their reference states. Method 4 again outperforms the other methods, with an MAE of 29.38 meV/atom, at least 10 meV/atom lower than the others. Method 2 has higher MAEs for both decomposition and formation energies (41.22 and 52.43 meV/atom, respectively) compared to other methods that freeze AtomRef during training, suggesting that a trainable AtomRef may lead to less accurate predictions in practice.

In the prediction of decomposition energies, we also observed that the uMLIP trained with Method 2 and Method 3 exhibited some failed ionic relaxations. Specifically, we found that in Method 2, 40 out of 34,927 relaxations, and in Method 3, 30 out of 34,927 relaxations, resulted in at least one atom being displaced more than 6 \AA\ away from its nearest neighbors, creating an unrealistic atomic configuration that triggered the failure of force field calculations. This is likely due to the unstable PES in the MLIP created by the large gradient updates in TL without shifting the reference energy (see Fig.~\ref{training}). In contrast, Method 4 -- TL with r$^2$SCAN AtomRef, significantly improves prediction accuracy in this complex task of predicting non-intrinsic properties.

\subsection{Scaling law on transfer learning}

\begin{figure}[htb]
\centering
\includegraphics[width=0.905\linewidth]{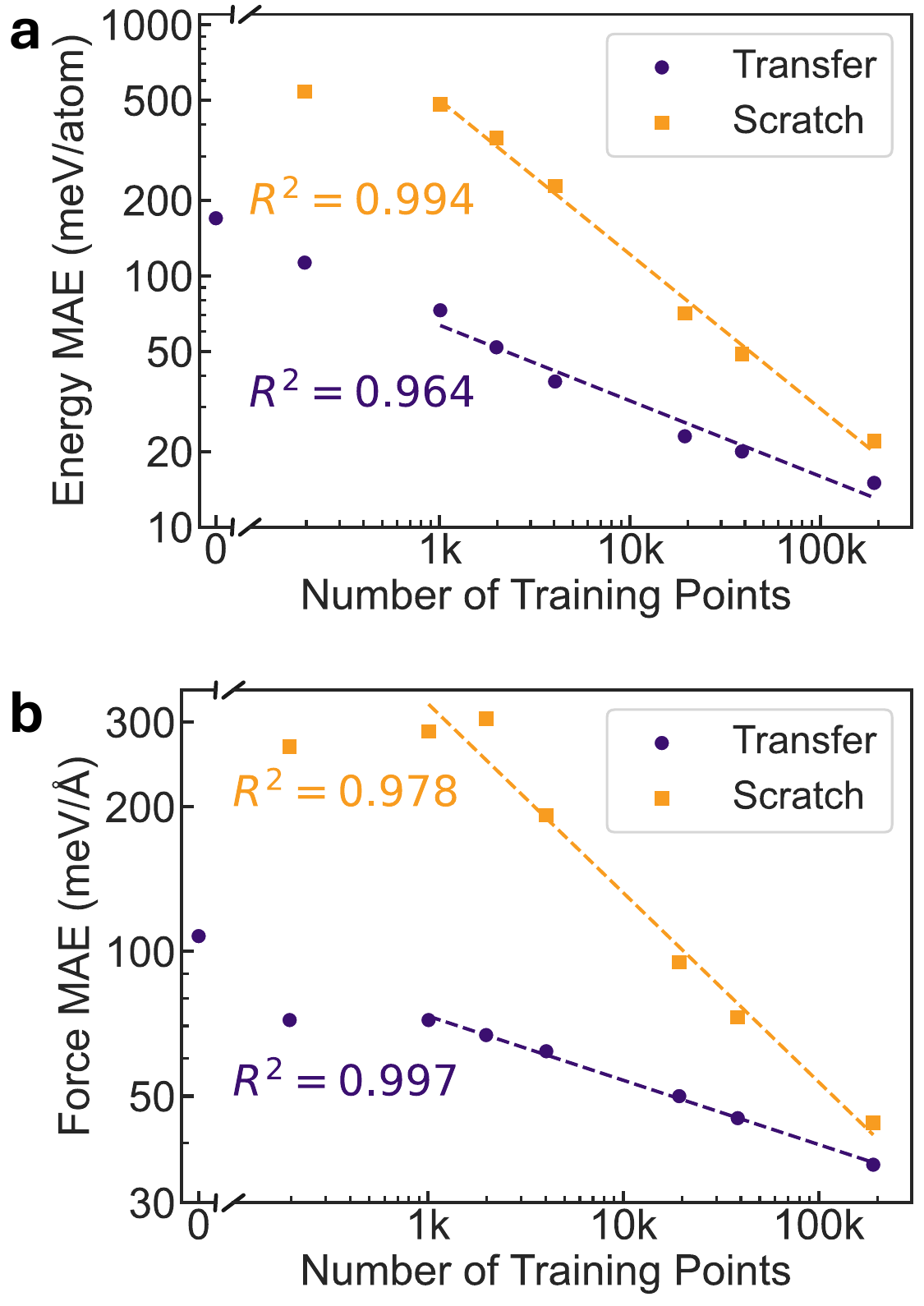}
\caption{\textbf{Scaling law on r$^2$SCAN data}. \textbf{a} Energy MAE and \textbf{b} Force MAE on the MP-r$^2$SCAN validation set using either Method 4, TL with r$^2$SCAN AtomRef (Transfer, blue) or Method 1, training from scratch (Scratch, orange) methods. Zero training points in Transfer refers to the performance of the GGA/GGA$+U$ pre-trained CHGNet with r$^2$SCAN AtomRef. Linear fits are applied for $x>1000$ to demonstrate the neural scaling law, and the coefficients of determination (R$^{2}$) are shown in the figures.}
\label{scaling_r2SCAN}
\end{figure}

To evaluate the data efficiency improvement of Method 4, we analyzed its scaling behavior on the MP-r$^2$SCAN dataset. The neural scaling laws suggest that model performance should improve steadily as the model size, dataset size, and amount of computing used for training are increased~\cite{Bahri_2024_scaling, Frey_2023, merchant2023scaling}. The performance is expected to follow a power-law relationship with each of these factors, provided the other two are not limiting. We benchmarked the energy and force MAEs on the validation set of MP-r$^2$SCAN using either Method 1 (Scratch) or Method 4 (Transfer). The resulting validation errors vs. training sizes are shown in Figure~\ref{scaling_r2SCAN}. For each curve in Fig.~\ref{scaling_r2SCAN}, we performed a linear regression starting from the data point corresponding to more than 1,000 training points on the x-axis, yielding the coefficient of determination (R$^{2}$) shown in the figures. The Linear fits demonstrate a linear scaling law behavior for both training from scratch (orange) and transfer learning (blue). The best-performing model for both energy and force predictions is obtained by Transfer, with an energy MAE of 15 meV/atom and a force MAE of 36 meV/\text{\AA}.

The superior data-efficiency of TL over training from scratch can be found by the reduced MAE of TL in Fig.~\ref{scaling_r2SCAN}. 
For energy MAE in Fig.~\ref{scaling_r2SCAN}a, the Scratch curve exhibits a log-log slope of -0.615 with an R$^{2}$ of 0.994, while the Transfer curve has a log-log slope of -0.301 with an R$^{2}$ of 0.964. 
For force MAE in Fig.~\ref{scaling_r2SCAN}b, the Scratch curve shows a log-log slope of -0.394 with an R$^{2}$ of 0.978, while the Transfer curve has a log-log slope of -0.134 with an R$^{2}$ of 0.997. 
The results indicate TL with merely 1K high-fidelity data points can outperform training from scratch on a high-fidelity dataset with more than 10K data points, marking more than 10-fold data efficiency gained from the GGA pre-training step.

Interestingly, we observe that the superior performance of Transfer over Scratch does not saturate even given the full-sized MP-r$^2$SCAN dataset of 0.24 million structures. Assuming the linear scaling trend of both Transfer and Scratch, the superior performance of Transfer will only be saturated after 719,996 training points for energy and 317,475 training points for force. This result indicates TL remains data-efficient even with close-to-million scale high-fidelity data points.

\section{Discussion and Summary}

The uMLIPs enable efficient predictions of energy across diverse chemical environments, facilitating large-scale simulations with near GGA-level accuracy. As the training of uMLIPs is migrating toward higher levels of DFT accuracy, optimal transferability strategies are needed. 
In this work, we investigated and benchmarked different transfer learning methods for uMLIPs with multi-fidelity datasets. We demonstrate that the scale of atomic reference energies varies significantly across different approximate density functionals, leading to the non-trivial choice of fine-tuning and TL approaches. We rationalized the importance of refitting the atomic reference energies when fine-tuning MLIPs across multi-fidelity datasets.

The energy quantity that matters for physical behavior is always referenced to some reference energies and not determined by total energies. For example, the cohesive energy is referenced to the energy of neutral, free atoms at infinite separation~\cite{kittel2018introduction}. The formation energy is referenced to the energy of constituent elemental unaries in their reference states (solid or gas phase)~\cite{xin2009point}, and decomposition energy is referenced to the energies of competing compounds in a given chemical space~\cite{bartel2020critical}. Consequently, the eV/atom scale shifts in total energy from GGA/GGA$+U$ to r$^2$SCAN do not lead to any changes in the physical interaction and behavior of materials. However, as energy is the training label for a ML model, the significant difference in the energy scales leads to challenges in the convergence of the TL. 

Essentially, by using energy referencing, one can modify the energy loss component in a model's loss function during TL. For a uMLIP with AtomRef, the general formula for the modified energy loss error of a structure's data is:
\begin{equation}
\begin{aligned}
    Loss^\text{Energy} = E_\text{label}^{\text{target}} & - (E_{\text{GNNs}}^{\text{source}} + \mathbf{c}_{\text{elem}} \cdot \mathbf{E}_{\text{elem}}^{\text{source}}) \\ 
    & - \mathbf{c}_{\text{elem}} \cdot (\mathbf{E}_{\text{ref}}^{\text{target}} 
    - \mathbf{E}_{\text{ref}}^{\text{source}}),
\end{aligned}
\end{equation}
where $E_\text{label}^{\text{target}}$ is the target energy training label, which is often obtained from high-fidelity calculations. $\mathbf{c}_{\text{elem}}$ is the composition row vector representing the number of each element in the structure. $\mathbf{E}_{\text{elem}}^{\text{source}}$ represents the AtomRef of the source dataset. $E_{\text{GNNs}}^{\text{source}}$ and $\mathbf{c}_{\text{elem}}\cdot \mathbf{E}_{\text{elem}}^{\text{source}}$ are the energy predictions of the GNN and AtomRef, which sum up to the energy prediction of the source uMLIP that has been pre-trained from a low-fidelity source dataset. $\mathbf{E}_{\text{ref}}^{\text{target}}$ and $\mathbf{E}_{\text{ref}}^{\text{source}}$ are the energy referencing parts of the two functionals, with dimensions $N_{\text{elem}} \times 1$, representing the reference energies of the structures. For cohesive energy, the reference energies are the energies of neutral free atoms at rest; for formation energy, they are the energies of unaries in their reference states. In our approach, they are also coming from the fitted AtomRefs. 

Energy referencing refers to replacing the AtomRef from $\mathbf{E}_{\text{elem}}^{\text{source}}$ to $(\mathbf{E}_{\text{elem}}^{\text{source}} + \mathbf{E}_{\text{ref}}^{\text{target}} - \mathbf{E}_{\text{ref}}^{\text{source}})$ before transferring a uMLIP to the target level. After energy referencing, the remaining contribution in the energy loss represents the differences in atomic interactions approximated by the source (GGA/GGA$+U$) versus the target (r$^2$SCAN), which is the relevant part of the energy that TL on GNNs aims to learn. 
Using AtomRef as $\mathbf{E}_{\text{ref}}$ is potentially better than referencing related to cohesive or formation energy, as AtomRef obtains atomic reference energies as statistical averages from all data in the dataset that covers a vast chemical space.

We attribute the effectiveness of using AtomRef as $\mathbf{E}_{\text{ref}}$ for cross-functional TL to two key factors. Firstly, the more than 10-fold improvement in correlation from 0.0917 to 0.9250 (see \hyperref[approach]{TL with different atomic reference energies}) significantly enhances the effectiveness of TL.  Secondly, refitting AtomRef ensures gradual adjustments of the model weights, and thus a more stable and reliable training process. Without refitting AtomRef, energy shifts cause substantial discrepancies between predicted and target energies, leading to very large prediction errors and high loss values initially. This, in turn, produces large gradients that cause excessive changes with the model weights, as illustrated in Fig.~\ref{training}a and b. 

According to Table~\ref{efsm_decomp_maes}, Method 4 (TL with r$^2$SCAN AtomRef) is shown to be most effective with the lowest energy MAE, consistent with the above rationalization of this approach. The higher prediction MAEs of Methods 2 (TL with trainable AtomRef) and 3 (TL with frozen AtomRef) compared to Method 4 — which integrates energy re-referencing with GNN-based TL — highlight the challenges of conventional TL without refitting AtomRef in uMLIPs. Methods 2 and 3 exhibit similar MAEs since they both begin with GGA/GGA$+U$ AtomRef, and the large energy shifts between r$^2$SCAN and GGA/GGA$+U$ cause poor correlation and excessive weight adjustments during early fine-tuning, driving model weights to suboptimal positions where they can become trapped. Notably, their predictions for forces, stresses, and magmoms are inferior to those of Method 1 (Training from scratch), which uses r$^2$SCAN data directly, free from GGA/GGA$+U$ influence. This underperformance is attributed to negative transfer~\cite{wang2019characterizing}, resulting from the weak correlation between source and target datasets during GNN-based TL.

As it is unlikely that one dataset will rule all of uMLIPs, a well-founded strategy to integrate diverse datasets, such as Materials Project~\cite{jain2013commentary}, Alexandria~\cite{ghahremanpour2018alexandria}, OQMD~\cite{saal2013materials}, AFLOWLIB~\cite{curtarolo2012aflow}, NOMAD~\cite{draxl2019nomad}, QM9~\cite{ramakrishnan2014quantum}, JARVIS~\cite{choudhary2020joint}, OC20~\cite{chanussot2021open}, OMat24~\cite{barroso2024open}, OCX24~\cite{abed2024open}, and MatPES~\cite{kaplan2025foundational}, will provide a promising avenue for leveraging the broad spectrum of available information and enable integration of future high quality data. Such integration will be helpful to address the data-originated issues in uMLIPs which are otherwise challenging to solve by only model architecture improvements~\cite{deng2025systematic}. Our scaling law analysis demonstrates the superior data efficiency gained from pre-training on large-scale low-fidelity dataset when migrating to high-fidelity ones.

As uMLIP-training is expected to transfer to higher quantum chemistry levels of theory, we also want to highlight the need to establish benchmark tests tailored to these computationally demanding quantum mechanical methods, such as \rrscan, coupled cluster methods (e.g., CCSD), and multi-reference approaches, as exemplified by our work on stability benchmarks using decomposition energy and formation energy predictions. Current uMLIP benchmarks such as Matbench Discovery~\cite{riebesell2023matbench} are mostly limited to GGA/GGA$+U$ tasks due to the dataset limits. We advocate for more comprehensive benchmarking frameworks that go beyond GGA/GGA$+U$ and potentially integrate evaluations such as kinetic properties and more complex material behavior to better assess models across different functionals.

In summary, by examining how atomic reference energies influence the performance of GGA/GGA$+U$ to \rrscan TL, we reiterate the importance of establishing correlations between multi-fidelity datasets so that they can benefit from TL. TL with refitting atomic reference energies yields a stable and reliable MLIP for energy, interatomic forces, and thermodynamic stability prediction. Our benchmark results and scaling law analysis show that refitting atomic energy is data-efficient and convinces fine-tuning uMLIPs to be a practical way for various downstream materials modeling tasks.

\section*{Acknowledgments}
This work was funded by the U.S. Department of Energy, Office of Science, Office of Basic Energy Sciences, Materials Sciences and Engineering Division under Contract No. DE-AC0205CH11231 (Materials Project program KC23MP). The work was also supported by the computational resources provided by the Extreme Science and Engineering Discovery Environment (XSEDE), supported by National Science Foundation grant number ACI1053575; the National Energy Research Scientific Computing Center (NERSC), a U.S. Department of Energy Office of Science User Facility located at Lawrence Berkeley National Laboratory; and the Swift Cluster resource provided by the National Renewable Energy Laboratory (NREL). The authors thank Luca Binci and Lauren Walters for valuable discussions.

\section*{Methods}
\label{sec:methods}

\textbf{Data preparation.} 
\label{subsec:data}The r$^2$SCAN Dataset, MP-r$^2$SCAN, is parsed from the Materials Project Database in March 2024. We collected all the r$^2$SCAN structure optimization and static task trajectories under each material ID that contain these tasks, and then following similar criteria as those used in creating the MPtrj Dataset:
(1) Final frame energies were limited to within 20 meV/atom of the primary task.  
(2) Structures missing energy, forces, or electronic convergence were excluded.  
(3) Structures with energies \textgreater~1eV/atom or \textless~10 meV/atom relative to Materials Project's ThermoDoc relaxed structures were filtered out to eliminate large energy differences resulting from variations in DFT calculation settings. 
(4) Duplicate structures were removed using pymatgen's StructureMatcher and energy matcher~\cite{ong2013python}. For all 4 TL models, we randomly split the MP-r$^2$SCAN dataset into training, validation, and test sets with an approximate ratio of 8:1:1 based on material IDs. The training set contains 27,943 material IDs with 190,560 structures; the validation set contains 3,492 material IDs with 23,888 structures; and the test set contains 3,492 material IDs with 23,799 structures. The energy, force, stress, and magmom prediction MAEs are based on the test set's 23,799 structures. The decomposition energy prediction MAE was reported on the test set. The formation energy prediction MAE was calculated on all 34,938 r$^2$SCAN material IDs in the Materials Project.

\textbf{Training scheme.} We kept most of the settings the same as the pre-trained CHGNet model, except for the following: we changed the fixed GGA/GGA$+U$ AtomRef of the model to r$^2$SCAN AtomRef; a Huber loss with energy, force stress and magmom loss ratio of 3:1:0.1:1 was used to train the model; we used a batch size of 64 and a learning rate of  \(10^{-3}\) that cosinely decays to  \(10^{-5}\) in 50 epochs.

\textbf{Feature importance.} 
\label{subsec:fea_impor}
To determine which elements contribute most to the formation energy differences between r$^2$SCAN and PBE/PBE$+U$ (discussed in Section \hyperref[subsec:functional_difference]{Energy differences across two functionals}), we used the attribute \texttt{feature\_importances\_} in \texttt{scikit-learn}'s \texttt{DecisionTreeRegressor}. 

The importance of each node on the decision tree can be calculated by (assuming only two child nodes (binary tree)):
\begin{equation}
n_{j} = w_j \sigma_j - w_{\text{left}(j)} \sigma_{\text{left}(j)} - w_{\text{right}(j)} \sigma_{\text{right}(j)}
\end{equation}
\(n_{j}\) represents the importance of node \(j\), \(w_j\) is the weighted number of samples reaching node \(j\), \(\sigma_j\) denotes the impurity value (here it is variance) of node \(j\), \(\text{left}(j)\) refers to the child node from the left split on node \(j\), and \(\text{right}(j)\) refers to the child node from the right split on node \(j\).

Feature importance is calculated by:
\begin{equation}
f_i = \frac{\sum_{j: \text{node } j \text{ splits on feature } i} n_{j}}{\sum_{k: \text{all nodes}} n_{k}}
\end{equation}
where \(f_i\) represents the importance of feature \(i\), and \(n_j\) represents the importance of node \(j\).

To obtain the normalized feature importance, each feature importance was divided by the total number of atoms of this element in the dataset and then multiplied by 9,000 for Fig.~\ref{data_analysis}c and 500 for Fig.~\ref{data_analysis}d to scale it back to the range of 0–1. Finally, it was visualized on the periodic table.

\section*{Data availability}
The MP-r$^{2}$SCAN dataset used to fine-tune CHGNet is available at 
\url{https://doi.org/10.6084/m9.figshare.28245650.v2} \cite{Huang2025MPr2SCAN}.

\section*{Acknowledgments}
This work was funded by the U.S. Department of Energy, Office of Science, Office of Basic Energy Sciences, Materials Sciences and Engineering Division under Contract No. DE-AC0205CH11231 (Materials Project program KC23MP). The work was also supported by the computational resources provided by the Extreme Science and Engineering Discovery Environment (XSEDE), supported by National Science Foundation grant number ACI1053575; the National Energy Research Scientific Computing Center (NERSC), a U.S. Department of Energy Office of Science User Facility located at Lawrence Berkeley National Laboratory; and the Swift Cluster resource provided by the National Renewable Energy Laboratory (NREL). The authors thank Luca Binci and Lauren N. Walters for valuable discussions.

\section*{Author Contributions}
B.D. and G.C. conceived the initial idea. X.H. performed Dataset Collection. X.H. benchmarked all the transfer learning methods. X.H. performed experiments on scaling law analysis. P.Z. and A.K. offered insights into the discussion of DFT functionals. B.D., K.P., and G.C. offered insights and guidance throughout the project. All authors contributed to discussions and approved the paper.

\section*{Competing interests}
The authors declare no competing interests.


\bibliography{ref.bib}

\end{document}